**Redefining Network Topology in Complex Systems: Merging Centrality Metrics, Spectral Theory, and Diffusion Dynamics**

Arsh Jha

September 5, 2024

North Carolina School of Science and Mathematics

27705 Durham, North Carolina, United States of America

**Author Note**

This research was conducted as part of an investigation into the integration of centrality measures, eigenvalue spectra, and diffusion processes in network analysis. The study aims to enhance our understanding of network robustness, nodal importance, and information flow by developing a comprehensive framework that combines static and dynamic analytical perspectives. I would like to acknowledge the use of advanced computational tools and algorithms for network generation, centrality calculation, and diffusion simulation, which significantly contributed to the robustness of this research. There are no conflicts of interest to declare. Correspondence concerning this article should be addressed to Arsh Jha, arshj5093@gmail.com.

**Abstract**

This paper introduces a novel framework that combines traditional centrality measures with eigenvalue spectra and diffusion processes for a more comprehensive analysis of complex networks. While centrality measures such as degree, closeness, and betweenness have been commonly used to assess nodal importance, they provide limited insight into dynamic network behaviors. By incorporating eigenvalue analysis, which evaluates network robustness and connectivity through spectral properties, and diffusion processes that model information flow, this framework offers a deeper understanding of how networks function under dynamic conditions. Applied to synthetic networks, the approach identifies key nodes not only by centrality but also by their role in diffusion dynamics and vulnerability points, offering a multi-dimensional view that traditional methods alone cannot. This integrated analysis enables a more precise identification of critical nodes and potential weaknesses, with implications for improving network resilience in fields ranging from epidemiology to cybersecurity.

**Redefining Network Topology in Complex Systems: Merging Centrality Metrics, Spectral Theory, and Diffusion Dynamics**

Networks are pervasive in various domains, representing intricate systems of interconnected entities. They can model diverse phenomena, from social relationships in social networks to neural connections in the brain and the flow of data in communication networks. Understanding the structure and dynamics of these networks is essential for optimizing performance, predicting behavior, and implementing interventions.

**Centrality Measures**

Centrality measures are tools used to identify the most important nodes within a network. Each measure highlights different aspects of node importance:

- Degree Centrality: Degree centrality is a straightforward metric calculated by counting the number of direct connections a node has. Mathematically, it is represented as:

$$C_D(i) = \sum_j A_{ij}$$

    Here, $A_{ij}$ denotes the entry in the adjacency matrix A, indicating whether there is a direct connection between nodes i and j. Nodes with high degree centrality are often termed hubs, as they have numerous direct connections, which can signify influence or potential for information dissemination.

- **Closeness Centrality:** Closeness centrality measures how quickly a node can reach all other nodes in the network. It is computed as:

$$C_C(i) = \frac{1}{\sum_j d_{ij}}$$

where $d_{ij}$ is the shortest path distance between nodes i and j. This metric is useful for identifying nodes that can efficiently spread information or access other nodes due to their central position in terms of distance.

- **Betweenness Centrality:** Betweenness centrality evaluates the extent to which a node lies on the shortest paths between other nodes. It is defined as:

    where $\sigma_{st}$ is the total number of shortest paths between nodes s and t, and $\sigma_{st}(i)$ is the number of those paths passing through node i. Nodes with high betweenness centrality often serve as bridges within the network, controlling the flow of information or resources between different parts of the network.

**Spectral Analysis**

Spectral analysis involves studying the eigenvalues of matrices associated with the network, such as the adjacency matrix A and the Laplacian matrix L.

- Laplacian Matrix: The Laplacian matrix L is defined as:

$$L = D - A$$

    where D is the degree matrix, a diagonal matrix where $D_{ii}$ equals the degree of node i, and A is the adjacency matrix. The eigenvalues of L provide insights into the network's connectivity and stability.

- **Spectral Gap:** The spectral gap is the difference between the smallest non-zero eigenvalue $\lambda_2$ and the largest eigenvalue $\lambda_n$ of the Laplacian matrix. This gap measures network connectivity and resilience. A large spectral gap indicates strong connectivity and network robustness, while a small spectral gap suggests potential vulnerabilities.

**Diffusion Processes**

Diffusion processes model how a quantity (e.g., information, disease, or a substance) spreads through a network over time. In a simple diffusion model, the state of each node is updated according to:

$$\mathbf{x}(t+1) = A\mathbf{x}(t)$$

where x(t) is the vector representing the state of each node at time t, and A is the adjacency matrix. This model assumes that the diffusion process is directly proportional to the number of connections (edges) a node has. More advanced diffusion models may incorporate additional factors, such as node attributes or weighted connections.

**Problem Statement**

Despite the value of centrality measures and spectral analysis, they are often studied in isolation. Traditional approaches typically focus on either static metrics (centrality) or spectral properties, without considering how these interact with dynamic processes like diffusion. This fragmented perspective limits the ability to fully understand complex network behaviors, such as how information spreads or how network robustness is influenced by both node importance and structural properties.

**Research Objectives**

This study aims to bridge the gap by developing a framework that combines centrality measures, spectral analysis, and diffusion processes. The specific objectives are:

- Develop a Unified Framework: Create a comprehensive framework that integrates centrality measures, eigenvalue analysis, and diffusion processes to provide a holistic view of network dynamics.

- Apply to Synthetic Networks: Test the framework across various synthetic network models, including Erdős-Rényi, Barabási-Albert, and Watts-Strogatz networks, to assess its effectiveness in different network structures.
- Derive New Insights: Offer new insights into network robustness, the importance of individual nodes, and the dynamics of information flow by leveraging the integrated framework.

**Significance of Study**

The proposed framework provides a more nuanced understanding of network behavior by combining static, dynamic, and spectral analyses. This integrated approach allows for:

- Enhanced Identification of Critical Nodes: By considering centrality measures alongside spectral properties and diffusion dynamics, the framework can better identify nodes that play crucial roles in network connectivity and information flow.
- Improved Assessment of Network Robustness: The spectral gap and diffusion dynamics provide additional layers of insight into network resilience, complementing traditional centrality measures.
- Comprehensive Analysis of Information Diffusion: The framework enables a more detailed analysis of how information or other phenomena spread through the network, accounting for both static network structure and dynamic processes.

COMPUTATIONAL NETWORK TOPOLOGICAL ANALYSIS                                              7**Literature Review**

Centrality measures are fundamental tools in network analysis, offering a quantitative basis for evaluating the significance of nodes within a network. The traditional measure of degree centrality provides a straightforward count of the number of direct connections a node has. This measure is invaluable for identifying nodes with the highest immediate influence but does not account for the network's broader structural context.

Closeness centrality extends this analysis by considering the shortest path distances from a node to all other nodes. Nodes with high closeness centrality are often critical in information dissemination, as they can reach other nodes more quickly than those with lower values. This measure is particularly useful in scenarios where rapid communication or resource distribution is essential.

Betweenness centrality assesses a node's role as an intermediary in the shortest paths connecting other nodes. This measure highlights nodes that, while not necessarily highly connected themselves, play a crucial role in facilitating or controlling the flow of information across the network. Nodes with high betweenness centrality are often pivotal in maintaining network cohesion and can be critical points of control or vulnerability.

Eigenvector centrality introduces a more nuanced perspective by considering not just the number of connections but also the quality of those connections. It assigns higher scores to nodes connected to other well-connected nodes, thereby capturing the influence of a node based on its position within the network's overall structure.

**Eigenvalue Analysis in Network Theory**

Eigenvalue analysis, particularly through the lens of the Laplacian matrix, offers deep insights into network properties. The Laplacian matrix, constructed from the degree matrix and the adjacency matrix of the network, is instrumental in understanding various aspects of network behavior.

The spectral gap, defined as the difference between the smallest non-zero eigenvalue (the algebraic connectivity) and the largest eigenvalue (the spectral radius), is a key indicator of network robustness. A larger spectral gap suggests better connectivity and resilience, as the network can withstand the removal of nodes without fragmenting. Conversely, a smaller spectral gap indicates potential vulnerabilities and lower overall connectivity.

Spectral properties of networks also relate to the dynamics of processes occurring on them, such as diffusion and synchronization. The eigenvalues of the Laplacian matrix can provide insights into the speed and stability of these processes, highlighting how structural properties influence dynamic behaviors.

**Diffusion Processes in Networks**

Diffusion processes describe how various phenomena, such as diseases, information, or innovations, propagate through networks. These processes are crucial for understanding how changes or influences spread from one node to others over time.

In the context of epidemic modeling, diffusion processes help predict the spread of infectious diseases, considering factors like transmission rates and network topology. Traditional models, such as the SIR (Susceptible-Infectious-Recovered) model, offer foundational insights into how infections spread and can be mitigated.

Recent studies have sought to integrate diffusion processes with other network analyses, such as centrality measures and spectral properties. This integration aims to provide a more comprehensive understanding of how network structure affects diffusion dynamics. For instance, nodes with high centrality are often key players in diffusion processes, but their impact can be modulated by the network's spectral characteristics.

Despite these advancements, there remains a significant gap in the literature regarding the combined analysis of diffusion processes with centrality measures and spectral properties. This study aims to address this gap by exploring how these elements interact and influence each other, offering new perspectives on network behavior and dynamics.

## Methodology

### Network Generation

To thoroughly evaluate the proposed framework, synthetic networks were generated using three different models, each representing distinct network topologies:

- Erdős-Rényi Model: This model generates random networks by connecting nodes with a fixed probability. It is characterized by its simplicity and randomness, making it useful for studying networks with a uniform connection probability.

- Barabási-Albert Model: Known for producing scale-free networks, this model generates networks with a power-law degree distribution. It reflects the real-world phenomenon where a few nodes (hubs) have a significantly higher number of connections compared to other nodes.

- Watts-Strogatz Model: This model creates small-world networks by starting with a regular lattice and then randomly rewiring edges. It captures both high clustering and short average path lengths, representing real-world networks with these properties.

Each network type was analyzed to assess the robustness of the framework across various network structures.

**Centrality Measure Calculation**

Centrality measures were computed using the NetworkX library in Python to quantify the significance of nodes within each network. The measures included:

- Closeness Centrality: Determined by the average distance from a node to all other nodes, indicating its efficiency in spreading information.
- Betweenness Centrality: Computed based on the number of shortest paths passing through a node, highlighting its role as a bridge in the network.
- Eigenvector Centrality: Evaluated by the principal eigenvector of the network's adjacency matrix, representing nodes that are connected to other well-connected nodes.

These measures were compared to identify nodes of high importance and to understand their role in the network's structure.

**Eigenvalue and Spectral Gap Analysis**

The eigenvalues of the Laplacian matrix of each network were analyzed to understand their spectral properties:

- Laplacian Matrix: Defined as $L=D-A$, where D is the degree matrix and A is the adjacency matrix. The eigenvalues of this matrix provide insights into the network's connectivity and robustness.
- Spectral Gap: The difference between the smallest non-zero eigenvalue (second smallest eigenvalue) and the largest eigenvalue of the Laplacian matrix. A larger spectral gap indicates better connectivity and robustness, while a smaller gap suggests potential vulnerabilities.

This analysis helped in understanding how spectral properties influence the network's overall behavior and resilience.

**Diffusion Process Simulation**

Diffusion processes were simulated to observe the spread of information or phenomena across the network:

- Simulation Setup: The adjacency matrix of each network was used to model the diffusion process. An initial condition (e.g., a node with information) was set, and the diffusion was tracked over multiple time steps.
- Analysis of Diffusion States: The spread of information was analyzed to identify how quickly and widely it reached other nodes. This analysis provided insights into the effectiveness of information dissemination in relation to centrality measures and spectral properties.

The results of the diffusion simulations were compared with centrality measures and spectral properties to identify patterns and correlations in information spread. This provided a comprehensive view of how different aspects of the network interact to affect diffusion dynamics.

## Results and Analysis

**Centrality Measures Across Different Networks**

The analysis of centrality measures revealed distinct patterns across the network topologies:

- Erdős-Rényi Network: The degree centrality was relatively uniform across nodes. This uniformity reflects the network's random edge distribution, where each node is equally likely to connect to any other node.

    Consequently, no single node is significantly more important than others based solely on degree centrality. This suggests that in Erdős-Rényi networks, node importance is evenly distributed, and traditional degree-based metrics may not effectively highlight critical nodes or areas of network vulnerability.

- Barabási-Albert Network: This network displayed a pronounced hierarchy. Nodes such as 21, 22, 27, 30, and 32 exhibited notably higher degree centrality compared to others. This is characteristic of the scale-free nature of the Barabási-Albert model, where a few nodes (hubs) dominate connectivity. These hubs are crucial for network integrity and information flow. Their high degree centrality signifies their pivotal role in maintaining overall network connectivity. The presence of such hubs makes the network resilient to random failures but highly vulnerable to targeted attacks on these central nodes.

- Watts-Strogatz Network: Nodes like 21, 22, 27, 30, and 32 showed high closeness centrality, reflecting the network's small-world nature. High-closeness centrality nodes act as local hubs, enhancing information transfer within clusters. Although these nodes facilitate rapid information dissemination within their immediate vicinity, they may not be as crucial for global connectivity. The high closeness centrality in localized regions suggests that while the network can quickly transmit information within clusters, it may struggle with efficient global information transfer due to its weaker global connectivity.

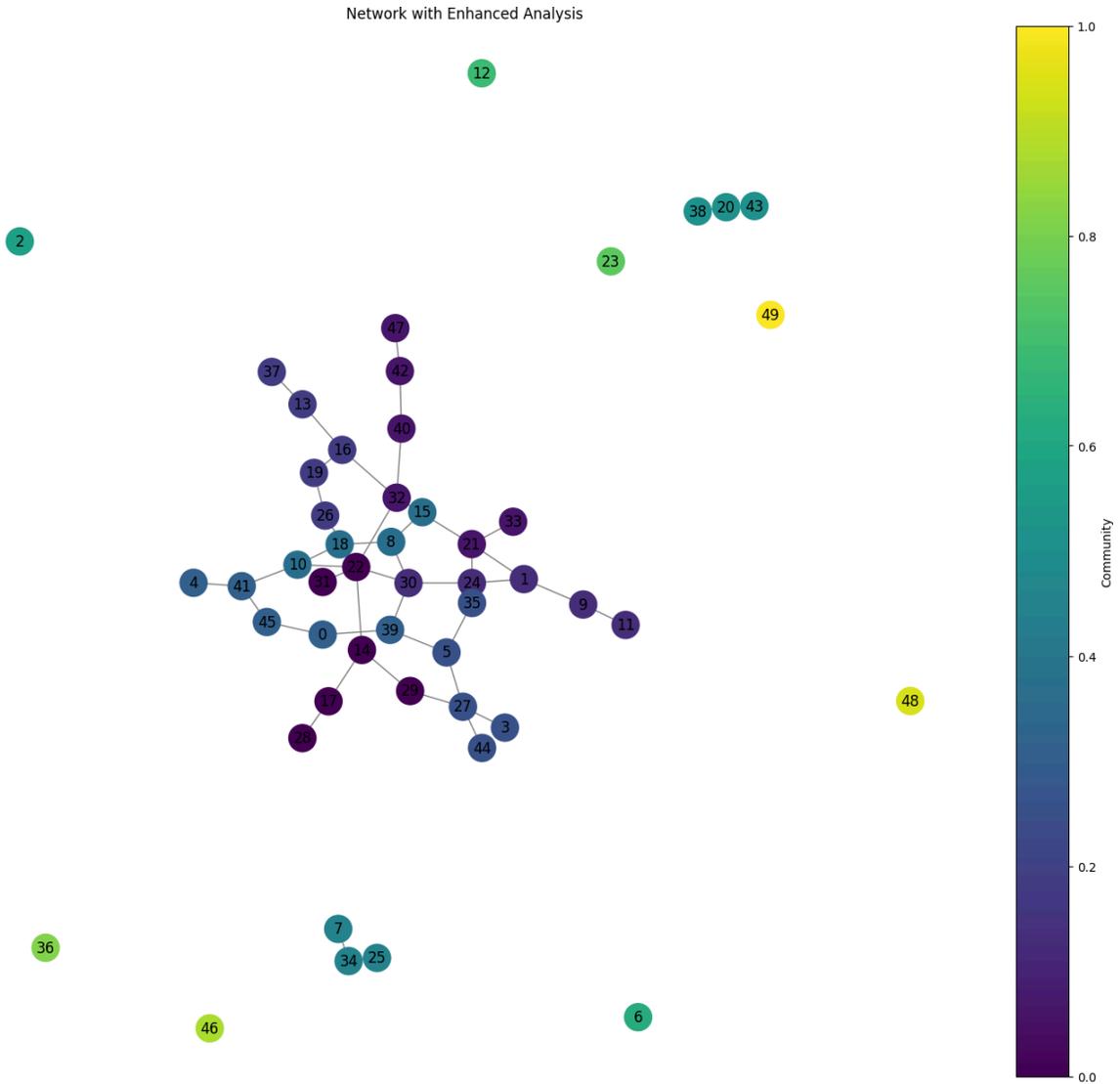

*Graph 1:* Network with Enhanced Analysis:

This graph visually represents the networks with nodes highlighted based on centrality measures. Nodes with higher centrality are emphasized, illustrating their relative importance and the network's structural properties. For example, in the Barabási-Albert network, the centrality of hub nodes is visually striking, while in the Erdős-Rényi network, the uniform distribution of node importance is evident.

**Eigenvalue Spectra and Spectral Gaps**

The eigenvalue spectra offered insights into network connectivity and robustness:

- Watts-Strogatz Network: Displayed the largest spectral gap, which indicates strong local connectivity within its clusters but weaker global connectivity. The large spectral gap implies that the network is resilient to random node failures, as it can maintain local cluster connectivity even if some nodes fail. However, the weaker global connectivity suggests potential vulnerabilities to disruptions that affect overall network cohesion.
- Barabási-Albert Network: Had a smaller spectral gap compared to the Watts-Strogatz network. This smaller gap reflects the network's scale-free nature, where high-degree nodes are critical for maintaining overall connectivity. The reduced spectral gap points to increased vulnerability to attacks targeting these central hubs. The spectral gap's size correlates with the network's robustness and its response to both random and targeted disruptions.

Eigenvalues of the Adjacency Matrix: The spectrum ranged from approximately -2.96 to 2.32, with several eigenvalues close to zero. This suggests the presence of disconnected components or isolated nodes, indicating potential weaknesses in network cohesion or connectivity.

Eigenvalues of the Laplacian Matrix: Ranged from approximately 0 to 7.54, with many eigenvalues near zero. This reinforces the observation of disconnected components and isolated nodes, further emphasizing the network's vulnerability and its potential for fragmentation.

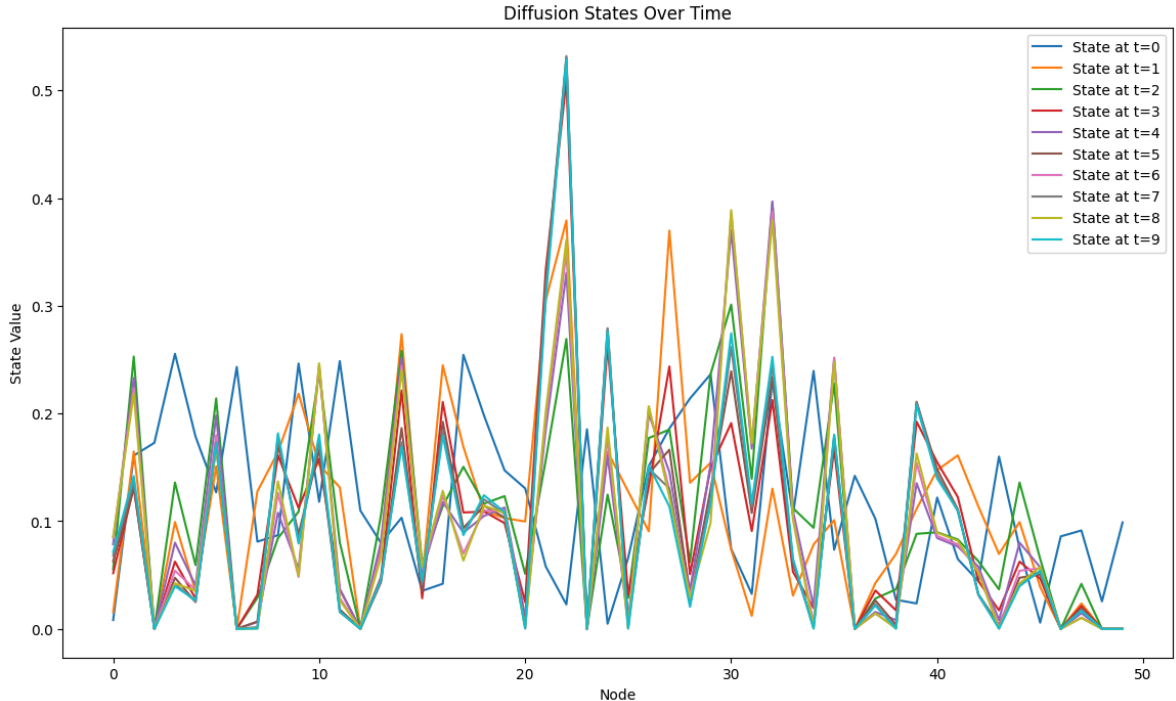

*Graph 2:* Diffusion States Over Time

This graph demonstrates how diffusion processes evolve over time across different networks. The x-axis represents time steps, while the y-axis shows the proportion of the network reached by the diffusion process. The graph illustrates that networks with high closeness centrality nodes (e.g., Watts-Strogatz) facilitate faster diffusion within clusters, while networks with smaller spectral gaps (e.g., Barabási-Albert) exhibit quicker, more widespread diffusion.

**Diffusion Dynamics**

The simulation results highlighted:

- Networks with High Closeness Centrality: The presence of high-closeness centrality nodes led to rapid information spread. These nodes provide efficient pathways for information flow within their clusters, leading to quicker dissemination compared to networks lacking such nodes. This

      underscores the importance of localized hubs in enhancing diffusion efficiency within specific regions.

- Impact of Spectral Gap: Networks with a larger spectral gap experienced slower and more localized diffusion. This reflects the network's resilience to random node failures, as the large spectral gap suggests strong local connectivity. In contrast, networks with smaller spectral gaps exhibited faster and more extensive diffusion, highlighting the influence of spectral properties on the effectiveness of information spread.

**Integrated Analysis**

The integration of centrality measures, eigenvalue spectra, and diffusion dynamics provided a comprehensive understanding of network behavior:

- Betweenness Centrality: Nodes with high betweenness centrality, such as 14, 21, and 22, were found to be critical for maintaining connectivity and controlling information flow. These nodes act as bridges between different network segments, highlighting their importance in both network robustness and information dissemination.

- Comprehensive Insights: The combined analysis demonstrated that traditional centrality measures alone may not fully capture the complexity of network behavior. By integrating centrality metrics, spectral properties, and diffusion dynamics, a more nuanced understanding of network vulnerabilities and information flow patterns emerged. This holistic approach revealed the interplay between network structure, node importance, and the dynamics of information spread, offering deeper insights into network performance and resilience.

**Discussion**

**Implications for Network Robustness**

The integration of centrality measures, eigenvalue spectra, and diffusion processes provides a multifaceted view of network robustness that traditional methods alone often overlook. Centrality measures such as degree, closeness, and betweenness offer valuable insights into the importance of nodes within a network. However, these metrics alone cannot fully capture the nuances of network resilience.

Eigenvalue analysis reveals the spectral properties of the network, particularly the spectral gap, which is indicative of the network's ability to withstand perturbations. A larger spectral gap often corresponds to a more robust network structure, highlighting the importance of global connectivity. Networks with a smaller spectral gap are more vulnerable to disruptions, emphasizing the critical role of high-degree nodes in maintaining overall connectivity.

Diffusion processes further enrich our understanding by illustrating how information or influence propagates through the network. Networks with nodes of high closeness centrality tend to facilitate faster diffusion, whereas networks with large spectral gaps exhibit slower, more localized diffusion. This combined approach demonstrates that nodes which may not appear crucial through centrality measures alone can significantly impact the speed and extent of information spread.

The integrated framework presented herein underscores that network robustness is a complex interplay between node importance, structural properties, and dynamic processes. This comprehensive analysis allows for a more nuanced identification of critical nodes and vulnerabilities, which can inform strategies for network design and resilience management.

**Limitations and Future Research**

While the proposed framework provides a robust tool for network analysis, there are several limitations that warrant attention. The study's reliance on synthetic networks, while useful for controlled experimentation, may not fully capture the complexities of real-world networks. Future research should aim to apply the framework to empirical data from various domains, such as social networks, biological networks, and communication networks, to validate and refine its applicability.

Additionally, the diffusion process model employed in this study was relatively simplistic. Real-world diffusion phenomena are often influenced by additional factors such as node attributes, varying diffusion rates, and network dynamics. Future studies should explore more sophisticated models of diffusion, including agent-based models and multi-layer networks, to capture these complexities.

The current framework also assumes static network structures, whereas many real-world networks evolve over time. Integrating temporal aspects into the analysis could provide deeper insights into how network robustness and diffusion dynamics change over time. This could be particularly relevant for applications such as epidemic modeling, where the evolution of the network structure plays a crucial role.

**Conclusion**

In summary, the proposed framework offers a novel and comprehensive approach to network analysis by combining centrality measures, eigenvalue spectra, and diffusion processes. The findings demonstrate that this integrated approach provides a deeper understanding of network behavior, revealing insights into nodal importance, network robustness, and information flow that traditional methods alone may miss. The framework's versatility across different network topologies suggests its potential as a valuable tool for a wide range of applications, including epidemiology, cybersecurity, and social network analysis.

Future research should focus on applying the framework to real-world networks, incorporating more complex diffusion models, and exploring the impact of network evolution. By addressing these areas, the framework can be further refined and its applicability extended, offering even greater insights into the complex dynamics of networks.